\documentclass[preprint,preprintnumbers,amsmath,amssymb,superscriptaddress]{revtex4}

\usepackage{graphicx}
\usepackage{dcolumn}
\usepackage{bm}

\begin{document}


\title{Experimental determination of the microscopic origin of magnetism in parent iron pnictides}

\author{D. Hsieh}
\affiliation{Joseph Henry Laboratories of Physics, Princeton
University, Princeton, NJ 08544, USA}
\author{Y. Xia}
\affiliation{Joseph Henry Laboratories of Physics, Princeton
University, Princeton, NJ 08544, USA}
\author{L. Wray}
\affiliation{Joseph Henry Laboratories of Physics, Princeton
University, Princeton, NJ 08544, USA}
\author{D. Qian}
\affiliation{Joseph Henry Laboratories of Physics, Princeton
University, Princeton, NJ 08544, USA}
\author{K.K. Gomes}
\affiliation{Joseph Henry Laboratories of Physics, Princeton
University, Princeton, NJ 08544, USA}
\author{A. Yazdani}
\affiliation{Joseph Henry Laboratories of Physics, Princeton
University, Princeton, NJ 08544, USA}
\author{G.F. Chen}
\affiliation{Beijing National Laboratory for Condensed Matter
Physics, Institute of Physics, Chinese Academy of Sciences, Beijing
100080, P.R. China}
\author{J.L. Luo}
\affiliation{Beijing National Laboratory for Condensed Matter
Physics, Institute of Physics, Chinese Academy of Sciences, Beijing
100080, P.R. China}
\author{N.L. Wang}
\affiliation{Beijing National Laboratory for Condensed Matter
Physics, Institute of Physics, Chinese Academy of Sciences, Beijing
100080, P.R. China}
\author{M.Z. Hasan}
\affiliation{Joseph Henry Laboratories of Physics, Princeton
University, Princeton, NJ 08544, USA} \affiliation{Princeton Center
for Complex Materials, Princeton University, Princeton, NJ 08544,
USA} \email{mzhasan@Princeton.edu}

\date{\today}


\maketitle

\textbf{Like high T$_c$ cuprates, the newly discovered iron based
superconductors lie in close proximity to a magnetically ordered
parent phase \cite{Kamihara, Hsu, cruz, Zhao, Sachdev}, suggesting its
superconductivity might be related to the mechanism driving its
magnetism. However, while the magnetic order in parent cuprates is
known to derive from a spin-spin interaction between electrons that
are localized due to strong Coulomb repulsion \cite{Lee},
a plethora of experiments including neutron scattering \cite{cruz, Zhao} have so far been unable to conclusively resolve whether a local moment Heisenberg description applies in parent iron based compounds \cite{Sachdev, Baskaran, Phillips, Si, Ma}, or whether
magnetism arises from a collective spin-density-wave (SDW) order
instability of an itinerant electron system like in chromium
\cite{Schafer}. These two alternatives can in principle be
distinguished by measuring the low energy momentum-resolved
bulk-representative electronic structure of the magnetically ordered phase. However,
whereas a single electronic band formed by one type of copper $d$-orbital contributes to the low energy electronic structure of the cuprates, the iron based compounds exhibit a complex multi-band manifold involving all five iron $d$-orbitals \cite{Ma, Mazin, Nekrasov, Singh}, making it challenging to experimentally track the dynamics of the electrons in the bands individually and identify the ones critical to the mechanism of magnetic and superconducting ordering. Using a combination of polarization dependent angle-resolved photoemission spectroscopy (ARPES) and scanning tunneling microscopy (STM), we have isolated the complete low-lying bulk representative electronic structure of magnetic SrFe$_2$As$_2$ with $d$-orbital symmetry specificity for the first time. Our results show that while multiple bands with different iron $d$-orbital character indeed contribute to charge transport, only one pair of bands with opposite mirror symmetries microscopically exhibit an itinerant SDW instability with energy scales on the order of 50 meV. The orbital resolved band topology below T$_{SDW}$ point uniquely to a nesting driven band hybridization mechanism of the observed antiferromagnetism in the iron pnictides, and is consistent with an unusual anisotropic nodal-density-wave state. In addition, these results place strong constraints on many theories of pnictide superconductivity that require a local moment quantum magnetism starting point.} \vspace{0.5cm}

Recent bulk measurements such as the quantum oscillation \cite{Sebastian} and optical \cite{Hu} measurements on SrFe$_2$As$_2$ support a spin-density-wave ground state where Fermi surface is directly involved, in sharp contrast to the ARPES studies \cite{Yang, Zhang, Liu2} that claim an absence of SDW gaps and argue for a local superexchange Heisenberg-type mechanism of magnetism. Since ARPES is a surface sensitive technique, it has been suggested that ARPES signals may not reflect bulk physics due to surface reconstruction or contamination effects. Here we show that bulk representative signals can in fact be isolated for the iron pnictides by comparing to STM surface topography results while combined with orbital-resolved polarized photoelectron signals.

\begin{figure*}
\includegraphics[scale=0.65,clip=true, viewport=0.0in 0in 9.5in 6.5in]{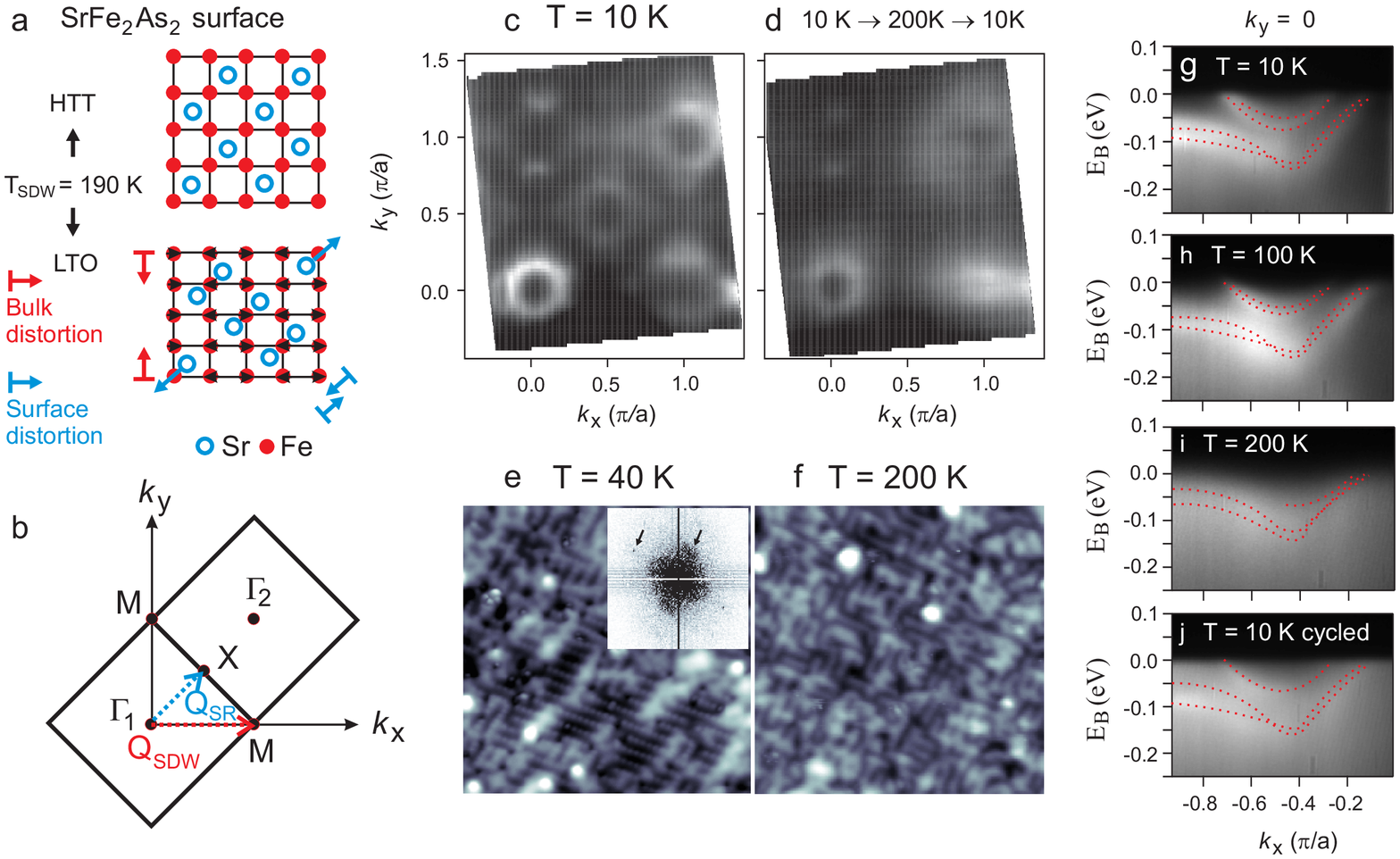}
\caption{\label{fig:SDW_Fig1} \textbf{Temperature dependent study of
near surface electronic structure of SrFe$_2$As$_2$.} \textbf{a,}
Schematic of the top-most layers of the crystal structure of
SrFe$_2$As$_2$ in the high temperature (non-magnetic) tetragonal
phase and in the low temperature (magnetically ordered) orthorhombic phase.
The Sr atoms are situated above the Fe plane and an intermediate As
layer has been omitted for clarity. The direction of the spins in
the orthorhombic phase are shown as black arrows on top of the red dots. \textbf{b,} Brillouin zones showing the wave vectors characterizing the bulk SDW ordering and surface reconstruction (SR). \textbf{c,} ARPES
intensity maps taken at $E_F$ following cleavage of the sample at 10
K and \textbf{d,} subsequent thermal cycling above T$_{M}$=190K, which
removes the pocket at X. \textbf{e,} STM topograph of the surface
(200 \AA) at 40 K. The inset is an fast Fourier transform profile of the topograph. The arrows point to the atomic peak and the 2 $\times$ 1 reconstruction of the surface which is also directly seen from the topography image. \textbf{f,} Topograph of the surface (200 \AA) taken after warming the sample up to 200 K, showing a significant lack of the long-range reconstruction. \textbf{g-j,} Temperature evolution of the
ARPES spectrum along the M-$\Gamma$ cut, which shows the
disappearance of two top bands near Fermi level ($\alpha_3$ and $\alpha_4$) above T$_{M}$ and reappearance of only on of them as the sample is cooled back below T$_{M}$.}
\end{figure*}

SrFe$_2$As$_2$ undergoes both a bulk structural and magnetic phase
transition at T$_{M}$ = 190 K, where the crystal structure changes
from tetragonal to orthorhombic and antiferromagnetic order with
wave vector \textbf{Q}$_{AF}$ = ($\pi$, 0) is achieved \cite{Zhao}.
STM studies have shown that low temperature cleaved surfaces of the
$A$Fe$_2$As$_2$ ($A$ = Ba, Sr, Ca) family consist of a nearly
complete $A$ layer sitting atop an As square lattice layer
\cite{Boyer, Yin}, and that the surface $A$ layer exhibits a
reconstruction with wave vector \textbf{Q}$_{SR}$ = ($\pi$/2,
$\pi$/2) (Fig.~\ref{fig:SDW_Fig1}\textbf{a},\textbf{b},\textbf{e}).
An ARPES intensity map taken at the Fermi level ($E_F$) of a 10 K
cleaved sample is displayed in Fig.~\ref{fig:SDW_Fig1}\textbf{c},
which shows features at $\Gamma$ = (0, 0) and M = ($\pi$, 0), (0,
$\pi$) in accordance with bulk band structure calculations \cite{Ma,
Dong, Ran, Singh, Nekrasov}. However there is an additional pocket
centered about X = ($\pi$/2, $\pi$/2) that does not appear in any
calculation. To show that this X pocket arises from a band folding
due to the surface reconstruction, we first note that its geometry
resembles that of the pocket enclosing $\Gamma$ and that it locates
exactly at $\Gamma+$\textbf{Q}$_{SR}$. Second, upon raising the
temperature above T$_{M}$, the surface reconstruction of Sr loses
its long-range coherence (Fig.~\ref{fig:SDW_Fig1}\textbf{f}),
concurrent with the disappearance of the ARPES feature at X
(Fig.~\ref{fig:SDW_Fig1}\textbf{d}). Although the ARPES features at
$\Gamma$ and M become less sharp after thermal cycling due to
surface contamination, they are largely insensitive to changes in
surface atomic arrangement, providing further evidence that these
states are bulk representative. Having isolated the bulk derived
states, we proceed to investigate their temperature dependence (see
SI for detailed study).
Figures~\ref{fig:SDW_Fig1}\textbf{g}-\textbf{j} show the evolution
of the electronic structure along the $\Gamma$-M direction through a
temperature cycle across T$_{M}$. While bands $\alpha_1$ and
$\alpha_2$ seem to simply broaden as the sample is heated, half of
the parabolic bands ($\alpha_3$ and $\alpha_4$) that exhibit mirror
symmetry about $k_x$ $\approx$ (0, -$\pi$/2) at low temperatures
disappear above T$_{M}$. Upon cooling back to 10 K, only the
$\alpha_3$ band reappears with much stronger intensity compared to
$\alpha_4$. Such behavior suggests that the $\alpha_3$ and
$\alpha_4$ bands may originate from bulk band folding due to
\textbf{Q}$_{AF}$, and may belong to different domains.

\vspace{0.5cm}

\begin{figure*}
\includegraphics[scale=0.7,clip=true, viewport=0.0in 0in 9.8in 4.8in]{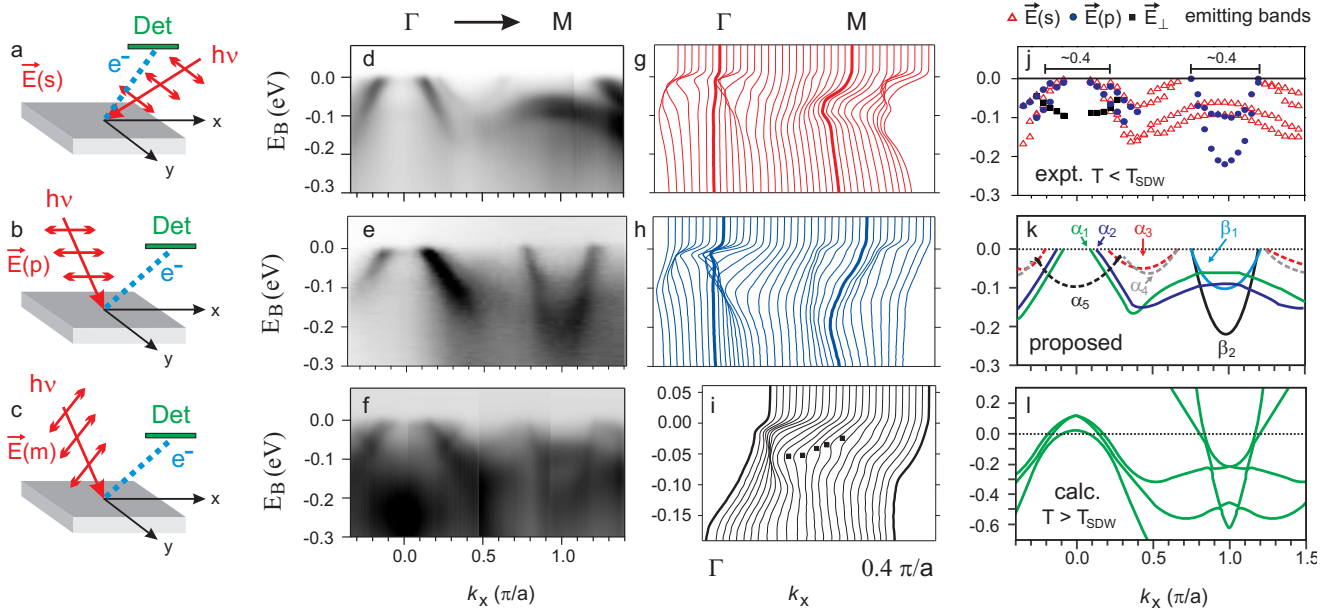}
\caption{\label{fig:SDW_Fig2} \textbf{Identification of band orbital
character through polarization dependent ARPES and comparison to LDA
calculation.} ARPES experimental configurations for \textbf{a,}
\textbf{E}$_s$, \textbf{b,} \textbf{E}$_p$ and \textbf{c,}
\textbf{E}$_m$ measurement geometries. \textbf{d-f,} ARPES spectra
of SrFe$_2$As$_2$ recorded at 10 K along the $\Gamma$-M cut, taken
under respective geometries. \textbf{g-i,} The corresponding energy
distribution curves. \textbf{j,} The complete low-lying band
structure in the magnetically ordered phase obtained by tracing the
peak positions of polarization dependent spectra. Bands that are
only visible under \textbf{E}$_s$ geometry have parity-odd orbital
symmetry, those that are only visible under \textbf{E}$_p$ have
parity-even orbital symmetry.
\textbf{k,} The proposed band connectivity derived from our
experimental data. \textbf{l,} A comparison with the LDA band structure of
SrFe$_2$As$_2$ in the non-magnetic phase \cite{Ma} shows that the
measured bands exhibit a large band dependent renormalization (note
energy scales), and suggests that bands $\alpha_3$, $\alpha_4$ and
$\alpha_5$ are related to the magnetically ordered phase since they are the additional bands after accounting for the surface reconstruction. These results also suggest that to understand the magnetic groundstate one needs to carry out systematic study of these bands.}
\end{figure*}

In order to determine the orbital character of the near-$E_F$
electronic bands, we show ARPES spectra along the $\Gamma$-M
($\parallel k_x$) cut taken with different incident photon
polarization directions to selectively excite states with different
orbital symmetry. Figures~\ref{fig:SDW_Fig2}\textbf{a,b,d,e} show
that changing the electric field direction from being oriented
perpendicular to (\textbf{E}$_s$) to being oriented parallel to
(\textbf{E}$_p$) the $\Gamma$-M direction results in the complete
suppression of one set of bands around M and the revelation of a new
non-overlapping set. Specifically, the $\alpha_3$ and $\alpha_4$
bands as well as the near M segment of bands $\alpha_1$ and
$\alpha_2$ are only visible under \textbf{E}$_s$ geometry (see
labels in fig.~\ref{fig:SDW_Fig2}\textbf{k}). Bands $\beta_1$ and
$\beta_2$, on the other hand, are only visible under \textbf{E}$_p$
geometry. Moreover, the near $\Gamma$ segments of bands $\alpha_1$
and $\alpha_2$, seen in figures~\ref{fig:SDW_Fig2}\textbf{d,e} as a
single dark band of intensity, appear polarization independent. When
the electric field vector has components along all three $x$, $y$
and $z$ directions (\textbf{E}$_m$), both sets of bands seen
independently under \textbf{E}$_p$ and \textbf{E}$_s$ geometries
become visible, and an additional electron-like band centered about
$\Gamma$ emerges (Fig.~\ref{fig:SDW_Fig2}\textbf{f}). Such a
dramatic polarization dependence can be explained by the spatial
symmetries of the near tetrahedrally coordinated Fe $3d$ complex,
which consist of nearly degenerate $t_{2g}$ triplet $d_{xy}$,
$d_{xz}$ and $d_{yz}$ closest to $E_F$ and a degenerate $e_g$
doublet $d_{x^2-y^2}$ and $d_{z^2}$ \cite{Calderon, Phillips, Si,
Nekrasov} (SI). Since $\Gamma$-M lies in a plane of mirror symmetry
(\textit{xz}-plane) in the crystal, electronic states that are even
(odd) under reflection with respect to this mirror plane can only be
excited by light with electric field polarization pointing in (out
of) the mirror plane \cite{Damascelli}. As the $d_{xz}$,
$d_{x^2-y^2}$ and $d_{z^2}$ orbitals have even symmetry, we conclude
that bands $\beta_1$ and $\beta_2$ have $d_{xz}$, $d_{x^2-y^2}$ or
$d_{z^2}$ character. Similarly, we conclude that bands $\alpha_3$,
$\alpha_4$ and the near M segments of bands $\alpha_1$ and
$\alpha_2$ have $d_{xy}$ or $d_{yz}$ character due to their odd
symmetry. The near $\Gamma$ segments of bands $\alpha_1$ and
$\alpha_2$ on the other hand are visible under all geometries, which
indicates that they are formed by some hybridization of odd and even
symmetric orbitals. The case for band $\alpha_5$ is less clear and
may involve orbitals with a larger out-of-plane extent such as
$d_{z^2}$ because a finite \textbf{E}$_z$ is required. Since the
ARPES intensity is proportional to $\langle
\textbf{E}\cdot\textbf{k} \rangle$ \cite{Damascelli}, the vanishing
of intensity near $\Gamma_n = (0,0,n2\pi)$ under \textbf{E}$_p$ and
\textbf{E}$_s$ geometries is expected. The highly $d$-like character
of all these bands is supported by first principles calculations
showing little weight of As $p$-orbitals near $E_F$ \cite{Cao}. Our
results show that polarization dependent ARPES must be employed to
exhaustively map the low energy bands of the iron pnictides, and
explains the difficulty of interpreting previous single polarization
ARPES studies \cite{Zhang, Yang, Lu}. The full experimental band
structure shown in figure~\ref{fig:SDW_Fig2}\textbf{j,k} shows
qualitative resemblance to LDA calculations
(Fig.~\ref{fig:SDW_Fig2}\textbf{l}) but exhibits a large band width
renormalization that ranges from ~2-3 times (it is $\sim$4-5 for
cuprates) depending on the band, similar to LaOFeP \cite{Lu}. While
strong $k_z$ dependence may explain this discrepancy, calculations
and experiment both show only very weak dispersion of all bands
along $k_z$ \cite{Borisenko, Nekrasov}. Instead, moderate electron
correlations may be at play. The experimental band structure reveals
a match between the $k_x$ extent of the odd symmetry $\alpha_3$
pockets around $\Gamma$ and the even symmetry $\beta_1$ and
$\beta_2$ pockets around M. To explore this possible nesting
instability, a full Fermi surface geometry is required.

\vspace{0.5cm}

\begin{figure*}
\includegraphics[scale=0.65,clip=true, viewport=0.0in 0in 11.0in 6.5in]{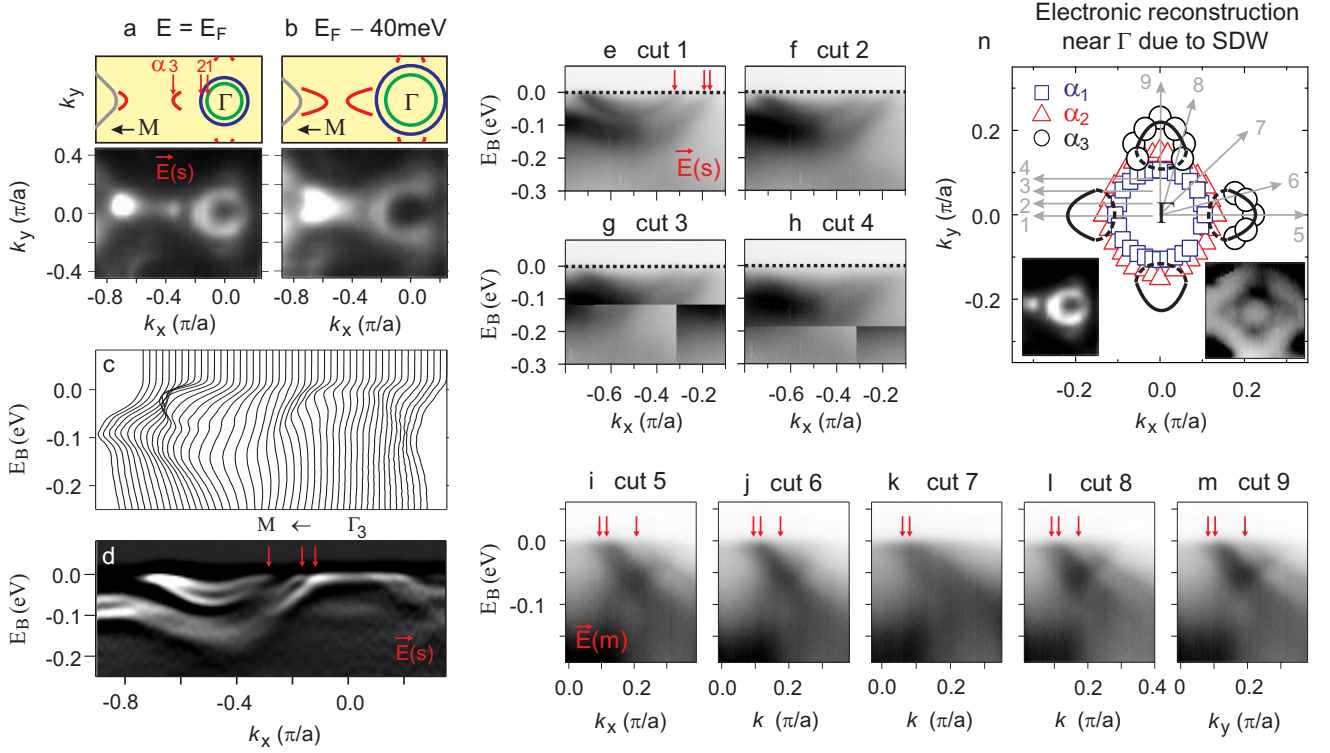}
\caption{\label{fig:SDW_Fig3} \textbf{Electronic reconstruction near
$\Gamma$ in the SDW magnetic phase.} \textbf{a,} ARPES intensity map
at $E_F$ and \textbf{b,} $E_F$ - 40 meV of SrFe$_2$As$_2$ at 10 K
around $\Gamma$ point taken with 40 eV photons in \textbf{E}$_s$
geometry. Data were collected in the third Brillouin zone but have
been mapped back to the first BZ for clarity. The schematics on top
of each image show the topology of the intensity distribution
observed. The red arrows point to positions of Fermi level crossings of
bands $\alpha_1$, $\alpha_2$ and $\alpha_3$ along the M-$\Gamma$
cut. \textbf{c,} Energy distribution curves along the M-$\Gamma$
direction and \textbf{d,} its corresponding second derivative image.
Bands $\alpha_1$, $\alpha_2$ and $\alpha_3$ are seen to cross $E_F$
at the $k_x$ positions marked by red arrows. \textbf{e-h,} ARPES
spectra taken using 30 eV photons along cuts 1 to 4. \textit{Band $\alpha_3$
gradually sinks completely below $E_F$ away from the $k_y$ = 0 line,
which indicates that $\alpha_3$ forms disconnected Fermi pockets
that do not enclose $\Gamma$ rather than a hole pocket that encloses
$\Gamma$.} \textbf{i-m,} ARPES spectra taken along cuts 5 to 9 in the
first BZ (panel \textbf{n}) using 35 eV photons under \textbf{E}$_m$
geometry.
\textbf{n,} Complete Fermi surface around $\Gamma$ obtained from ARPES cuts similar to those shown in panels (a-m). The size of the symbols
shown in \textbf{n} reflect the degree of k$_z$ dispersion exhibited by the pockets. Bottom left inset shows overall topology of the SDW reconstructed Fermi surface at $\Gamma$ to be contrasted with the topology of the Fermi surface observed at X due to band folding from surface reconstruction. The observed Fermi surface pockets around $\Gamma$ can be understood within a nesting mechanism shown in Fig.-5}
\end{figure*}

ARPES intensity maps near $\Gamma$ were collected using
\textbf{E}$_s$ geometry to capture all the bands (see
Fig.~\ref{fig:SDW_Fig2}) and are shown in
figures~\ref{fig:SDW_Fig3}\textbf{a,b} at different binding
energies. Due to some $k_z$ dependence of the bands \cite{Borisenko,
Ma, Nekrasov}, we found that the splitting between the $\alpha_i$
bands is largest in the third BZ using 40 eV photons
(Fig.~\ref{fig:SDW_Fig3}\textbf{c,d}), allowing for better
resolution of the individual Fermi surface pieces. Bands $\alpha_1$
and $\alpha_2$ form two concentric hole Fermi surfaces enclosing
$\Gamma$, as evidenced by their growth in size with increasing
binding energy. Due to a drastic weakening of the $\alpha_3$ band
intensity away from the $\Gamma$-M line at h$\nu$ = 40 eV, we had to
use 30 eV photons to follow its dispersion away from $\Gamma$-M,
which compromises the resolution of the $\alpha_1$ and $\alpha_2$
bands. Figures~\ref{fig:SDW_Fig3}\textbf{e-h} show a series of
constant $k_y$ cuts near $\Gamma$ with $k_y$ moving progressively
away from zero. For $k_y$ exceeding approximately 0.05 $\pi$, the
$\alpha_3$ band is seen to drop entirely below $E_F$, indicating
that its forms disconnected Fermi pockets that do not enclose
$\Gamma$ rather than a hole pocket that does enclose it as is the
case for bands $\alpha_1$ and $\alpha_2$. In order to test for
possible anisotropy of bands $\alpha_1$ through $\alpha_4$ between
the $k_x$ and $k_y$ directions owing to the bulk lattice distortion,
we performed azimuthal cuts around the $\Gamma$ FS in the first BZ
by rotating the sample so as to keep the scattering geometry
identical (\textbf{E}$_m$) for each cut
Figs~\ref{fig:SDW_Fig3}\textbf{i-m}. As the polar angle $\theta$ is
rotated away from $k_x$ ($\theta = 0^{\circ}$), bands $\alpha_1$ and
$\alpha_2$ remain gapless and retain a nearly constant Fermi wave
vector $k_F$ up to $\theta = 90^{\circ}$. Bands $\alpha_3$ and
$\alpha_4$ however weaken in intensity as $\theta$ approaches
45$^{\circ}$ ($\Gamma$-X line), disappear entirely at 45$^{\circ}$
(Fig.~\ref{fig:SDW_Fig3}\textbf{k}), and then grow more intense
again as $\theta$ approaches 90$^{\circ}$. The $k_F$ location of
band $\alpha_3$ traces the outer segment of a petal shaped FS
(Fig.~\ref{fig:SDW_Fig3}\textbf{n}) while the $\alpha_5$ band never
exhibits any intensity near $E_F$, suggesting that it does not
contribute to any FS. Altogether, these results show that there
exists a largely four-fold rotation symmetric FS near $\Gamma$ in
the magnetically ordered phase, which consists of two concentric
hole pockets enclosing $\Gamma$ and petal shaped hole pockets that
do not enclose $\Gamma$ (Fig.~\ref{fig:SDW_Fig3}\textbf{n}). These
$\alpha_1$ $\alpha_2$ and $\alpha_3$ pockets have sizes, ignoring their k$_z$ dependence, of around 1.6(5)\%, 2.7(9)\% and 1.0(4)\% of the BZ respectively, which are consistent with the two sizes 0.52\% and 1.38\% seen in quantum
oscillation measurements \cite{Sebastian}.

\vspace{0.5cm}

\begin{figure*}
\includegraphics[scale=0.6,clip=true, viewport=0.0in -0.3in 11.0in 8.3in]{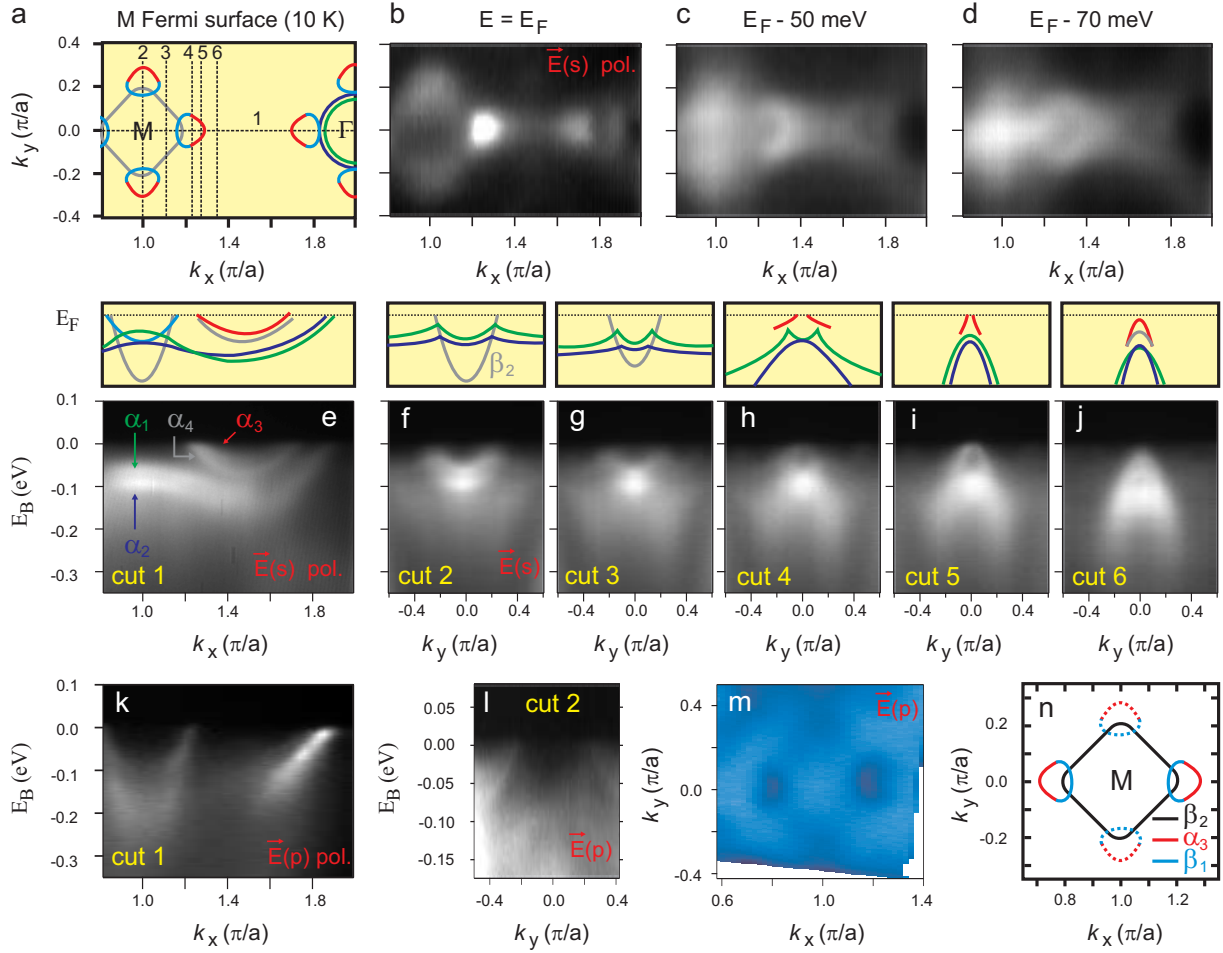}
\caption{\label{fig:SDW_Fig4} \textbf{Electronic reconstruction near
M in the SDW magnetic phase.} \textbf{a,} Schematic of the low
temperature Fermi surface topology of SrFe$_2$As$_2$ and the scan-line cut
positions. ARPES intensity map at \textbf{b,} $E_F$, \textbf{c,}
$E_F$ - 50 meV and \textbf{d,} $E_F$ - 70 meV taken with 30 eV
photons in \textbf{E}$_s$ geometry at 10 K around M point, which
show a diamond-shaped electron like Fermi surface enclosing M and
the outer segment of a hole like Fermi pocket to its side.
\textbf{e-j,} ARPES spectra taken along cut directions 1 to 6 taken
in \textbf{E}$_s$ geometry. The schematics on top of each image show
the band dispersions and the formation of the electron and hole like
Fermi surfaces near M. \textbf{k,} ARPES spectrum along cut 1 and
\textbf{l,} cut 2 taken with 30 eV photons under \textbf{E}$_p$
geometry. Band dispersion along cut 2 shows that the small hole
pockets also exist along the $k_y$ direction, though with far weaker
intensity. \textbf{m,} ARPES intensity map at $E_F$ taken under
\textbf{E}$_p$ geometry, which reveals the inner segment of the hole
pocket next to M. The fact that the diamond-shaped Fermi surface
enclosing M is still visible suggests that the $\beta_2$ band has a
mixed odd and even orbital symmetry away from the $\Gamma$-M line.
\textbf{n,} A summary of experimentally determined Fermi surface around M is presented. The dashed lines indicate that the hole pockets along $k_y$ are observed to be far weaker than those along $k_x$.}
\end{figure*}

The ARPES intensity distribution at $E_F$ around M taken in
\textbf{E}$_s$ geometry (Fig.~\ref{fig:SDW_Fig4}\textbf{a,b})
reveals a diffuse diamond-shaped ring of intensity enclosing M and
the outer segments of two bright petals of intensity located at
either side of M along $k_x$ (only one such spot located at
\textbf{k} $\approx$ ($1.3\pi, 0$) is shown for clarity). Probing
these features at higher binding energies
(figs~\ref{fig:SDW_Fig4}\textbf{c,d}) shows firstly that the
diamond-shaped FS exhibits electron-like character while the petal
FS exhibits hole-like character, and secondly that these two FS
pieces are disconnected, which suggests that they originate from
different bands. By comparing the shape of the petal shaped feature
near M and the petal shaped feature near $\Gamma$ (located at
\textbf{k} $\approx$ ($1.7\pi, 0$)) at all binding energies, it is
clear that they are completely reflection symmetric about $k_x$ =
1.5$\pi$, the BZ boundary of the magnetic unit cell, which indicates
that these FS features result from a band folding due to magnetic
ordering. In order to study the band origin of these Fermi surfaces,
we show band dispersion spectra along several momentum space cuts
near M taken in \textbf{E}$_s$ geometry. A cut along $k_x$ through M
(Fig.~\ref{fig:SDW_Fig4}\textbf{e}) shows that the bright petal FS
arises from the Fermi crossings of the $\alpha_3$ band, while bands
$\alpha_1$ and $\alpha_2$ do not contribute any intensity close to
$E_F$ near M along $k_x$. A cut along $k_y$ through M
(Fig.~\ref{fig:SDW_Fig4}\textbf{f}) shows that bands $\alpha_1$ and
$\alpha_2$ both exhibit an M-shaped dispersion whose maxima do not
reach $E_F$ (see SI for detailed analysis), and continue to sink
further below $E_F$ as $k_x$ moves away from $\pi$
(Fig.~\ref{fig:SDW_Fig4}\textbf{g-j}), eventually giving way to the
appearance of the hole-like $\alpha_3$ and $\alpha_4$ bands near
$E_F$. We attribute the intensity of the diamond-shaped Fermi
surface to the $\beta_2$ band, which is observed to be present but
very weak in figure~\ref{fig:SDW_Fig4}\textbf{f} (see SI for clear
distinction of $\beta_2$ band), and likely explains the low
temperature negative Hall coefficient \cite{Chen}. Further support
comes from studying the ARPES intensity distribution at $E_F$ around
M using \textbf{E}$_p$ geometry (Fig.~\ref{fig:SDW_Fig4}\textbf{m}),
under which the $\beta_1$ and $\beta_2$ bands appear most intense
(Fig.~\ref{fig:SDW_Fig4}\textbf{k}). Here, it is the inner segment
rather than the outer segment of the hole petals that is strongest,
marking an abrupt change of orbital character between the inner and
outer halves of the petal FS. As the Fermi crossing position of the
inner segment along $k_x$ matches well with that measured for
$\beta_1$ (Fig.~\ref{fig:SDW_Fig2}\textbf{k}), we conclude that the
hole petal is composed of distinct halves deriving from $\alpha_3$
and $\beta_1$. A cut along $k_y$ through M under \textbf{E}$_p$
geometry (Fig.~\ref{fig:SDW_Fig4}\textbf{l}) shows a band dispersion
very similar to the cut along $k_x$ under \textbf{E}$_s$ geometry
(Fig.~\ref{fig:SDW_Fig4}\textbf{e}), which provides evidence that
the petal FSs also exist along the $k_y$ direction, though with far
weaker intensity. Altogether, these data reveal a FS near M in the
magnetically ordered phase of SrFe$_2$As$_2$ that consists of a
diamond-shaped electron pocket enclosing M approximately 3.8(8)\% of
the BZ in size, and four smaller petal shaped hole pockets that do
not enclose M (Fig.~\ref{fig:SDW_Fig4}\textbf{n}), which are very
similar in shape and size to those observed around $\Gamma$.

\vspace{0.5cm}

\begin{figure*}
\includegraphics[scale=0.65,clip=true, viewport=0.0in 0in 11.0in 6.5in]{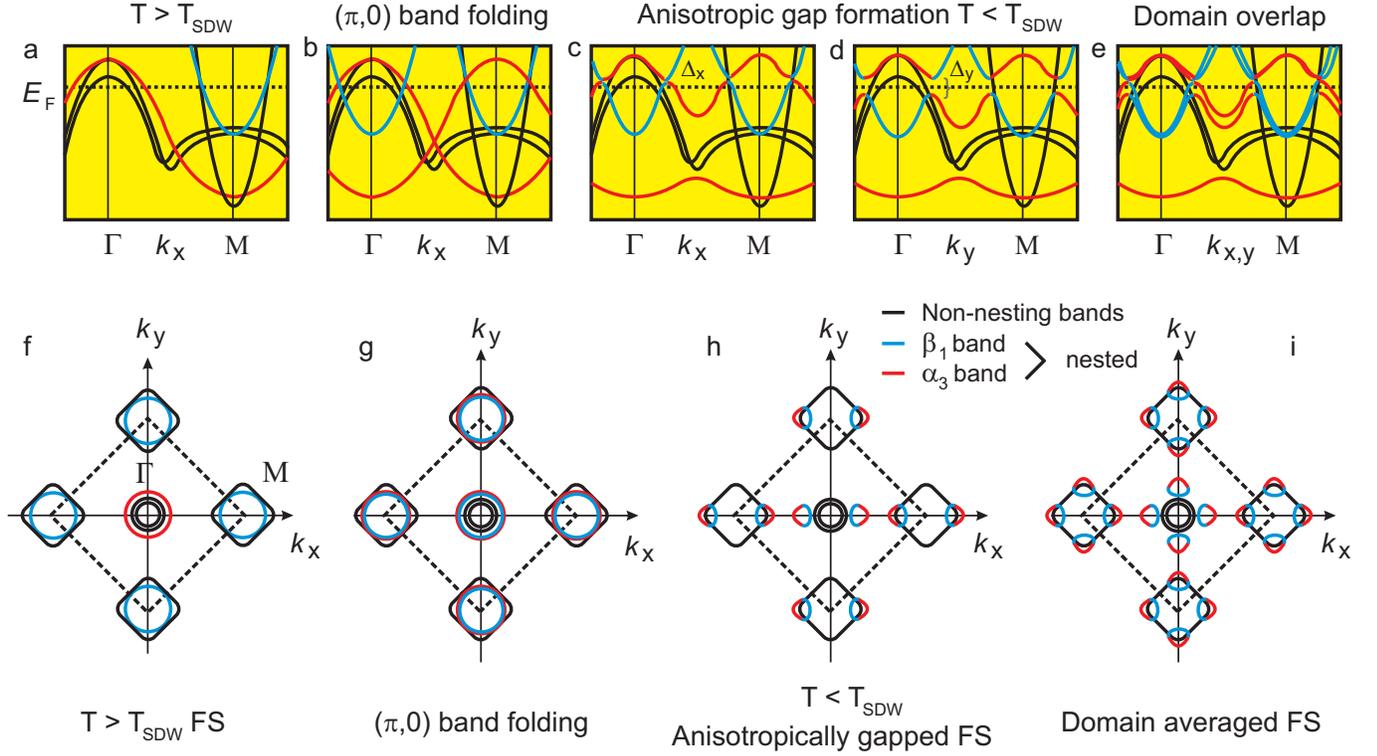}
\caption{\label{fig:SDW_Fig5} \textbf{Signature of orbital dependent
band nesting and an anisotropically gapped Fermi surface.}
\textbf{a,} Schematic of the proposed band structure of
SrFe$_2$As$_2$ in the high temperature non-magnetic phase and
\textbf{f,} its associated Fermi surface. The bands $\alpha_3$ and
$\beta_1$ that are nested by \textbf{Q}$_{SDW}$ are shown in red and
blue respectively. \textbf{b-d,} show the band folding that takes
place under \textbf{Q}$_{SDW}$ = ($\pi$, 0) and a $k_x$-$k_y$
anisotropic gap opening that occurs following hybridization of bands
$\alpha_3$ and $\beta_1$. The rest of the bands (black lines) do not
interact. \textbf{g-h,} Fermi surface resulting from anisotropic gap
formation necessarily consists of petal shaped hole pockets made up
of inner and outer segments with different orbital symmetry.
\textbf{e,} Domains of both \textbf{Q}$_{SDW}$ = ($\pi$, 0) and (0,
$\pi$) ordering in the sample give rise to the observed topology of the ARPES band structure along $\Gamma$-M and the shape and intensity distributions of the observed ARPES Fermi surface \textbf{i}. The schematic in (\textbf{i}) is based on the raw data presented in Fig.-3 and -4 and consistent with a recent theoretical proposal \cite{Ran}.}
\end{figure*}

The geometry of the SrFe$_2$As$_2$ FS and the momentum dependence of
its orbital character point clearly to a nesting driven SDW ground
state, which can be understood starting from the high temperature
LDA band structure (Fig.~\ref{fig:SDW_Fig5}\textbf{a,f}). Our ARPES
measurements suggest that only the $\alpha_3$ and $\beta_1$ bands
have nearly matching $k_F$ and thus have a significant ($\pi$, 0)
nesting instability. Therefore in a \textbf{Q}$_{SDW}$ = ($\pi$, 0)
SDW ordered phase, only the $\alpha_3$ and $\beta_1$ bands fold onto
one another (Fig.~\ref{fig:SDW_Fig5}\textbf{b}). In a conventional
SDW system where two Fermi surfaces are perfectly nested, the folded
bands hybridize and acquire an energy gap $\Delta$ over the entire
BZ, resulting in the complete destruction of any FS in the SDW phase
\cite{Gruner}. However, the presence of the petal hole pockets
indicates that this cannot be the case in SrFe$_2$As$_2$, and
suggests that either i) the high temperature $\alpha_3$ and
$\beta_1$ Fermi surfaces only partially nest due to a mismatch in
shape, or ii) the entirety of the $\alpha_3$ and $\beta_1$ Fermi
surfaces are nested but gap formation takes place anisotropically
($\Delta_x \neq \Delta_y$) around it. Since the former scenario
cannot explain the presence of both $\alpha_3$ and $\alpha_4$ bands,
we pursue the latter case. Anisotropic gap formation around the
nested regions as shown in figures~\ref{fig:SDW_Fig5}\textbf{c,d}
results in a two-fold symmetric FS with petal-like hole pockets that
have the exact dual orbital character as we have observed using
ARPES (Fig.~\ref{fig:SDW_Fig5}\textbf{h}). Such a generic two-fold
symmetric FS has recently been predicted by a 5 band model
\cite{Ran} and a first principles study \cite{Dong}. The former
study also claims that an SDW gap is not allowed to open along $k_x$
($\Delta_x$ = 0) due to the particular mirror symmetries of the
system, and thus the ground state is an intrinsically nodal SDW
\cite{Ran}. The fact that we observe a FS with four-fold rotation
symmetry rather than two-fold is very likely due to the coexistence
of domains with both \textbf{Q}$_{SDW}$ = ($\pi$, 0) and
\textbf{Q}$_{SDW}$ = (0, $\pi$)
(Fig.~\ref{fig:SDW_Fig5}\textbf{e,i}), as the photon beam used in
ARPES is much larger than the typical size of a single magnetic
domain, which explains the presence and temperature dependence of
both $\alpha_3$ and $\alpha_4$ bands
(Fig.~\ref{fig:SDW_Fig1}\textbf{g-j}). Although a four-fold
symmetric Fermi surface can in principle be realized in the 5 band
model \cite{Ran}, it requires a very particular combination of
interaction strengths and is therefore highly unlikely. The measured
gap $\Delta_y$ (Fig.~\ref{fig:SDW_Fig2}\textbf{h}) is roughly
consistent with weak coupling mean-field theory, which predicts
$\Delta$ = 3.53 $k_{B}$T$_{SDW}$ = 58 meV.
The gap size along the \textbf{Q}$_{SDW}$ direction ($\Delta_x$) on the other hand is
unclear since it lies above $E_F$.

In summary, our polarization dependent ARPES measurements, guided by STM results, reveal the full ground state electronic structure of SrFe$_2$As$_2$ for the first time, strongly pointing to a novel orbital selective anisotropically gapped SDW state driven by moderately correlated itinerant electrons. It is likely that as doping creates a size
mismatch between the $\alpha_3$ and $\beta_1$ pockets in the non-magnetic phase \cite{Borisenko, Wray}, nesting is weakened and a competing superconducting phase emerges, which is distinct from the doped Mott insulator approach that is taken for the
cuprates\cite{Lee} (Note that Ref-\cite{Borisenko} does not orbital-resolve the band topology). Residual ($\pi$,0) nesting fluctuations in the superconducting phase are likely bosonic modes contributing to Cooper pairing in doped iron pnictides. The experimental methods demonstrated here can be systematically applied to explore the electron behavior over the full phase diagram of pnictides and other multiband correlated electron systems.

\vspace{1cm}

We thank A. Pasupathy and A. Pushp for assistance with STM measurements. This work is supported by U.S. DOE/BES.

\begin{center}
\textbf{Methods}
\end{center}

Single crystals of SrFe$_2$As$_2$ were prepared using the methods
described in \cite{Hu}. ARPES measurements were carried out at the
Advanced Light Source beamlines 10.0.1 and 12.0.1 in order to
exploit multiple scattering geometries. Photon energies of 30 to 40
eV photons were used resulting in energy resolutions better than 15
meV and angular resolutions better than of the 1\% of the Brillouin
zone. Single crystalline samples of SrFe$_2$As$_2$ (T$_{SDW}$ = 190
K) were used for this study, which were cleaved at 10 K in chamber
pressures better than 3 $\times$ 10$^{11}$ torr unless otherwise
stated.

\newpage

\pagebreak

\end{document}